\documentclass[12pt]{article}
\usepackage{aas_macros}
\usepackage{epsfig}
\usepackage{amssymb}

\usepackage[svgnames,pdftex,hyperref]{xcolor}
\usepackage[pdftex]{hyperref}
\hypersetup{colorlinks=True,linkcolor=Black,citecolor=Black,
urlcolor=Black,pagebackref,pdftitle={The Zwicky Transient Facility}
,pdfauthor={Eric Bellm}}
\usepackage[numbers,square]{natbib}

\renewenvironment{thebibliography}[1]{%
  \begin{oldthebibliography}{#1}%
    \setlength{\parskip}{0ex}%
    \setlength{\itemsep}{0ex}%
}%
{%
  \end{oldthebibliography}%
}

\def\beq{\begin{equation}}
\def\eeq#1{\label{#1}\end{equation}}
\def\eeqn{\end{equation}}

\def\beqa{\begin{eqnarray}}
\def\eeqa#1{\label{#1}\end{eqnarray}}
\def\eeqan{\end{eqnarray}}

\let\bar=\overbar

\def\Dslash{\not{\hbox{\kern-4pt $D$}}}
\def\dslash{\not{\hbox{\kern-2pt $\del$}}}

\def\msb{{\bar{\ssstyle M \kern -1pt S}}}

\def\Title#1{\begin{center} {\Large {\bf #1} } \end{center}}

\begin{document}

\Title{The Zwicky Transient Facility}

\bigskip\bigskip


\begin{raggedright}  

{\it Eric Bellm\index{Bellm, E.}\\
Cahill Center for Astronomy and Astrophysics\\
California Institute of Technology
Pasadena, CA, USA}
\bigskip\bigskip
\end{raggedright}

\section{Introduction}

The Zwicky Transient Facility (ZTF; P.I. Shri Kulkarni) 
is a next-generation optical synoptic
survey that builds on the experience and infrastructure of the Palomar
Transient Factory (PTF) \citep{Law:09:PTFOverview,Rau:09:PTFScience}.  
Using a new
47\,deg$^2$ survey camera, ZTF will survey more than an order of 
magnitude faster than PTF to discover rare transients and variables.

PTF (and its successor survey, the Intermediate Palomar Transient Factory,
or iPTF) have conducted a transient-focused optical time-domain survey.
PTF uses a 7.26 deg$^2$ camera 
on the Palomar 48-inch Oschin
Schmidt telescope (P48) to survey the dynamic night sky in Mould-$R$ and SDSS
$g'$ bands.  
Followup photometry and spectroscopy are provided by the 60- and 200-inch
telescopes at Palomar and by other collaboration resources around the
world.

PTF's moderate-depth, followup-focused survey has yielded many notable
successes.  However, addressing leading-edge scientific questions (Section
\ref{sec:science}) 
requires a capability to survey at high cadence while maintaining
wide areal coverage.  Current facilities are inadequate for this purpose,
but a straightforward upgrade of the PTF survey camera provides this
capability while maintaining much of PTF's demonstrably productive hardware
and software infrastructure.  ZTF will provide the best characterization of
the bright to moderate-depth ($m \lesssim 21$) transient and variable sky
and pave the way for LSST's deeper survey.

\section{Survey Design}  \label{sec:design}

The traditional measure of \'{e}tendue (collecting area $\times$ solid angle)
is insufficient for characterizing the performance of time-domain
surveys \citep{Tonry:11:ATLAS}.  
It relates most closely to the speed at which an instrument
achieves a given coadded depth.  Time domain surveys are often interested
in the detection rate for specific classes of transients (e.g., Type Ia
SNe or Tidal Disruption Events).  
These detection rates are a function of the intrinsic rate,
brightness, and timescale of the transient; the cadence of the survey; and
the spatial volume surveyed in each cadence period.
For variability science, the utility of time series data depends on 
the limiting magnitude,
the photometric precision, the total number of observations, the cadence, and 
the bandpass(es) of the data.  

This wide range of survey parameter space indicates the difficulty of
optimizing a generic time-domain survey for a wide range of science goals.  
(It also
suggests that specialized surveys will continue to be productive into the era
of large time-domain facilities.)  
In consequence, single figures of merit are imperfect predictors of the
performance of a time-domain survey, as much depends on the specifics of
the chosen survey strategy in addition to the raw capabilities of the
camera and telescope.  However, optimization metrics are required to guide
design studies and cost trades.

Building on the PTF heritage, we have chosen to optimize the ZTF camera design 
for the study of explosive transients.  For any camera realization, we may
trade survey cadence against the sky area covered per survey snapshot.
We therefore seek to maximize the volumetric survey rate ($\dot{V}$), 
defined as the
spatial volume within which a transient of specified absolute magnitude
(e.g, $M = -19$) could be detected at $5\sigma$, divided by the total time
per exposure including readout and slew times.  With appropriate choice of
cadence, $\dot{V}$ should be proportional to the transient detection rate.
It implicitly incorporates the field of view of the camera, its limiting
magnitude (which in turn includes the image quality, sky background, 
telescope and filter throughput, and read noise), and overheads
\citep[c.f.][]{Tonry:11:ATLAS}.

Notably, specifying the overhead between exposures implies an optimal
exposure time to maximize $\dot{V}$.  Exposures that are too short lead to
an inefficient duty cycle, while exposures that are too long lead to
smaller snapshot volumes, as the loss of areal covered is not offset
by the increase in depth. 

Guided by these considerations, our design for the ZTF survey camera
(Section \ref{sec:camera}) maximizes the camera field of view, maintains
PTF's moderate image quality and depth, and minimizes the overhead between
exposures and the number of filters.  

\section{The ZTF Camera} \label{sec:camera}

The 7.26\,deg$^2$ field of view provided by the 
CFHT12k camera \citep{Rahmer:08:PTFCamera}
currently used by PTF only covers a fraction of the
available $\sim$47\,deg$^2$ focal surface of the P48.  By constructing a
new camera that fills the focal surface with CCDs, we obtain a
6.5 times larger field of view.  
Modern readout electronics will reduce the overhead
between exposures as well, providing a net improvement in survey speed 
of more than an
order of magnitude relative to PTF.  This speed boost will enable a
transformative survey capable of simultaneously maintaining the 
high cadence and wide areal coverage needed to find rare, fast, and young
transients.

The focal surface of the Schmidt telescope is curved, and during the
Palomar Sky Surveys the photographic plates were physically bent on a
mandrel to conform to this focal surface.  The PTF camera achieves
acceptable image quality (median 2'' FWHM in $R$) with a flat CCD focal
plane, an optically powered dewar window, and flat filters.  However, scaling a
comparable design to the full ZTF focal plane produces unacceptable image
quality.

We have developed an optical design that maintains PTF's image quality over
the entire available field of view.  An additional zero-power optic (to be
fabricated from an existing blank) placed in front of the existing achromatic
doublet Schmidt corrector provides a minor adjustment (10\%) to its aspheric
coefficient.  A faceted CCD focal plane and individual field flattener
lenses placed over each CCD correct the residual field curvature.
An optically powered window maintains vacuum in the dewar.  The optical
design supports exchangeable flat filters, or the filter coatings may be
deposited onto the field flatteners mounted over each CCD.

Improved yields for wafer-scale CCDs make large focal planes 
increasingly affordable.  ZTF will use 16 e2v 6k$\times$6k devices with 15
micron pixels.  At 1''/pixel, the pixel scale is identical to that of PTF
and adequately samples the median 2'' image quality.  The moderate pixel
scale also mitigates the data volume.  Six CCDs have been fabricated and
delivered as of this writing.
At 1 MHz readout, read time will be 10\,sec; we require 15\,sec net
overhead between exposures to allow for slew and settling.
With these shorter overheads, 30\,sec exposures are optimal in maximizing
$\dot{V}$.
A compact dewar design minimizes mass and beam obstruction.  

\begin{table}
\begin{centering}
\begin{tabular}{|l|r|r|}
\hline
Specification & PTF & ZTF \\
\hline
Active Area & 7.26 deg$^2$ & 47 deg$^2$ \\
Exposure Time & 60 sec & 30 sec \\
Readout Time & 36 sec & 10 sec \\
Median Time Between Exposures & 46 sec & 15 sec \\
Median Image Quality ($R$ band) & 2.0'' FWHM & 2.0'' FWHM \\
Median Single-visit Depth ($R$ band) & 20.7 & 20.4 \\
Yearly Exposures per Field ($3\pi$) & 19 & 290 \\
Areal Survey Rate & 247 deg$^2$/hr & 3760 deg$^2$/hr \\
Volumetric Survey Rate ($M=-19$) & $2.8\times10^3$ Mpc$^3$/s &
$3.0\times10^4$ Mpc$^3$/s \\
\hline
\end{tabular}
\caption{{\small Comparison of the PTF and ZTF cameras and
survey performance metrics. Yearly exposures assume a hypothetical
uniform $3\pi$ survey.}
\label{tab:ptf_vs_ztf}}
\end{centering}
\end{table}

Table \ref{tab:ptf_vs_ztf} compares the performance of the ZTF survey
camera to that of PTF.

\section{Selected Science Goals} \label{sec:science}

\subsection{Young SNe}

Observations of SNe within the first 24 hours of explosion reveal key 
information about their progenitors and environments.  Early photometric
observations of SNe Ia constrain the radius of the progenitor and
can distinguish single- and double-degenerate
scenarios \citep{Kasen:10:EarlyIaBinarity}.  In core-collapse 
SNe, early observations probe the poorly-measured physics of shock
breakout and shock heating \citep{Nakar:10:ShockBreakoutLCs}.
Early-time ``flash'' spectroscopy of core-collapse SNe within hours of the
explosion can directly measure
the properties of the circumstellar medium and reveal the final stages of stellar 
evolution before the explosion \citep{GalYam:14:FlashSpectroscopy}.

Detecting, discovering, and following up young transients in a single night
requires finely honed pipelines, procedures, and collaboration.  The
PTF and iPTF collaborations have demonstrated
the ability to obtain these time-critical measurements on several
occasions \citep{GalYam:11:PTFRapidIIp}.
However, the total number of young SNe in the PTF
datastream is limited by the survey camera: obtaining the few-hour cadence
observations needed to detect young SNe limits the survey to a much smaller
area of sky.  With ZTF's wider, faster camera, the collaboration will be
able
to systematically study a true sample of SN progenitors rather than an
isolated handful: we can detect twelve times more
SNe with ZTF at any chosen cadence.  In a high-cadence survey, 
ZTF will detect one SN within 24 hours of its explosion \textit{every night}.

\subsection{Fast-decaying transients}

While PTF, CRTS, and Pan-STARRS1 have occasionally 
observed at relatively high cadences
(images separated by less than a few hours), the correspondingly small 
areal coverage permitted by their survey cameras has limited the 
detection of fast transients to M-dwarf flares \citep{Berger:13:PS1FastTransients}.  ZTF's
order-of-magnitude increase in survey speed will place much tighter
constraints on the existence of fast-decaying explosive transients,
exceeding published limits on areal exposure in less than 
one week of observations.

One intriguing event, PTF11agg \citep{Cenko:13:PTF11agg}, highlights the
potential of ZTF in this area.  Discovered by PTF during high-cadence
monitoring of the Beehive Cluster for variable star studies, PTF11agg
declined by almost two magnitudes over several hours.  While its properties
are consistent with an optical afterglow of a gamma-ray burst (GRB), there
was no high-energy trigger from wide-field gamma-ray monitor.  This raises
the possibility that PTF11agg represents a new class of event, a
baryon-loaded ``dirty fireball'' that would not show MeV emission.
The inferred rate of such events would be
about twenty times the GRB rate.  

With ZTF's faster survey speed, we expect to detect more than 20
PTF11agg-like events per year, as well as a handful of classical GRB orphan
optical afterglows.  These measurements will place important constraints on the
opening angles of GRB jets as well as the diversity of relativistic stellar
explosions.

\subsection{Gravitational Wave Counterparts}

Beginning in 2015, advanced gravitational wave (GW) interferometers
will begin operations.  They are expected to detect
the first GW signals from binary neutron star mergers.  Detecting the
electromagnetic counterparts to these events will provide vital physical
information, including independent distance estimates and information about
the merger progenitors and host galaxy.  The mergers are predicted to
produce optical counterparts, whether from afterglows of short-hard
gamma-ray bursts or ``kilonovae'' powered by r-process
nucleosynthesis
\citep{LP98,Kulkarni:05:Macronovae,Metzger:2010,Kasen:13:KilonovaeOpacity}.

Unfortunately, the earliest GW detections will be very poorly localized,
with error boxes of hundreds of square degrees 
with only two detectors and improving to tens
of square degrees as more interferometers come online.
Detecting a rapidly-decaying optical transient with unknown brightness
in this large sky area is a monumental challenge.  Success
will require a well-tested technical stack, including all-sky reference images,
fast and reliable image differencing, a complete local galaxy catalog to
prioritize followup, and the ability to obtain rapid spectroscopy \citep{Kasliwal:13:LIGOLoc}.
iPTF has proven this approach by successfully using its transient pipeline
to localize the 
afterglows of Fermi-detected gamma-ray bursts within 70 square degrees
\citep{Singer:13:iPTF13bxl}.  ZTF's wider field will be vital for achieving
the same success with the larger search areas and fainter counterparts
provided by GW detections.

\subsection{Variability Science}

The repeated observations provided by PTF and other surveys 
have built an increasingly
valuable photometric variability catalog.
Single-filter time variability information may be used to identify and 
classify variable stars \citep{Richards:12:ASASCatalog}, 
identify binary systems, and measure
the mass of the supermassive black holes in AGN systems
\citep{Kelly:09:AGNVariability}.
Variable stars may be used to trace Galactic structure and identify dwarf
galaxies \citep{Drake:13:TidalStream,Sesar:13:OrphanStream}, 
thereby mapping the gravitational potential of the Milky Way and
testing predictions of $\Lambda$CDM cosmology
\citep{Bullock:05:DisruptedHalos}. 
Photometric variability may even predict stellar parameters, including
effective temperature, surface gravity, and metallicity
\citep{Miller:14:PredictingStellarParameters}.

ZTF's greater survey speed will provide an unprecedented variability
catalog.  If observations are spread evenly over the entire visible Northern
sky, we will obtain nearly 300 observations per field each year.  
CRTS currently provides the most uniform photometric variability coverage.
ZTF will provide a larger number of observations as well as improved
cadence and depth, enabling a wide range of variability science on sources
accessible to moderate-aperture telescopes and advancing community
involvement in advance of LSST's deeper survey.

\bigskip
E.~B. is grateful for useful conversations with Shri Kulkarni,
Tom Prince, Richard Dekany, Roger Smith, Jason Surace, Eran Ofek, Mansi
Kasliwal, Branimir Sesar, and Paul Groot.

\bibliographystyle{yahapj}
\bibliography{ptf} 

\end{document}